\documentclass{llncs}
\usepackage[citestyle=alphabetic,style=alphabetic,urldate=long]{biblatex}
\bibliography{kbib/kwarc}
\usepackage{latexml}
\title{\LaTeXML\ 2012 -- A Year of \LaTeXML}
\author{Deyan Ginev\inst{1} \and Bruce~R.~Miller\inst{2}}
\institute{Computer Science, Jacobs University Bremen, Germany
 \and National Institute of Standards and Technology, Gaithersburg, MD, USA}
\date{\today}

\begin{document}
\maketitle
\begin{abstract}
\LaTeXML, a \TeX\ to XML converter, is being used in a wide range
of MKM applications.
In this paper, we present a progress report for the 2012 calendar year.
Noteworthy enhancements include: increased coverage such as Wikipedia syntax;
enhanced capabilities such as embeddable JavaScript and CSS resources
and RDFa support; a web service for remote processing via web-sockets;
along with general accuracy and reliability improvements.
The outlook for an 0.8.0 release in mid-2013 is also discussed. 
\end{abstract}

\section{Introduction}

\LaTeXML\ \cite{Miller:latexml:online} is a \TeX\ to XML converter, bringing the well-known authoring syntax
of \TeX\ and \LaTeX\ to the world of XML. Not a new face in the MKM crowd, \LaTeXML\
has been adopted in a wide range of MKM applications. Originally designed to
support the development of NIST's Digital Library of Mathematical Functions (DLMF),
it is now employed in publishing frameworks, authoring suites and for the preparation
 of a number of large-scale \TeX\ corpora.

In this paper, we present a progress report for the 2012 calendar year of \LaTeXML's
 master and development branches. In 2012, the \LaTeXML\ Subversion repository saw 30\% of the total project commits since 2006. 

Currently, the two authors maintain a developer and master branch of \LaTeXML, respectively. The main branch contains all mature features of \LaTeXML. 

\section{Main Development Trunk}\label{sec:trunk}
\LaTeXML's processing model can be broken down into two phases:
the basic conversion transforms the \TeX/\LaTeX\ markup into
a \LaTeX-like XML schema; a post-processing phase converts that XML
into the target format, usually some format in the HTML family.
The following sections highlight the progress made
in support for these areas.

\subsection{Document Conversion}\label{sec:conversion}
There has been a great deal of general progress in \LaTeXML's
processing: the fidelity of \TeX\ and \LaTeX\ simulation is much improved;
the set of control sequences covered is more complete.
The I/O code has been reorganized to more closely track \TeX's behavior
and to use a more consistent path searching logic.
It also provides opportunities for more security hardening,
while allowing flexibility regarding the data sources,
needed by the planned web-services.
Together these changes allow the direct processing of many more
`raw' style files directly from the \TeX\ installation
(i.e., not requiring a specific \LaTeXML\ binding).  This mechanism is,
in fact, now used for loading input encoding definitions and
multi-language support (\texttt{babel}).
Additionally, it provides a better infrastructure for sTeX.

The support for colors and graphics has been enhanced,
with a more complete color model that captures the capabilities of
the \texttt{xcolor} package and a move towards generation of native SVG \cite{W3C03:SVG1.1}.
A summer student, Silviu Oprea, now at Oxford,
developed a remarkable draft implementation supporting
the conversion of \texttt{pgf} and \texttt{tikz} graphics markup into SVG;
this code will be integrated into the 0.8 release.

Native support for RDFa has been added to the schema,
along with an optional package, \texttt{lxRDFa}, allowing
the embedding of the semantic annotations within the \TeX\ document.
Various other \LaTeX\ packages have also been implemented:
\texttt{cancel}, \texttt{epigraph}.
Additionally, the \texttt{texvc} package provides for the
emulation of the \texttt{texvc} program used by Wikipedia
for processing math markup; this allows \LaTeXML\ to be used
to generate MathML from the existing wiki markup.

\subsection{Document Post-Processing}\label{sec:post}

The conversion of the internal math representation
to common external formats such as MathML and OpenMath
has been improved. In particular, the framework fully supports
parallel math markup with cross-referencing between
the alternative formats. Thus presentation and content
MathML can be enclosed within a \texttt{m:semantics} element,
with the corresponding \texttt{m:mi} and \texttt{m:ci} tokens
connected to each other via \texttt{id} and \texttt{xref} attributes.

The evolution of MathML version 3 has also been tracked,
as well as the current trends in implementations.
Thus, we have shifted towards generating SMP (Supplemental Multilingual Plane,
or Plane 1) Unicode and avoiding the \texttt{m:mfenced} element.
Content MathML generation has been improved, particularly
to cover the common (with \LaTeXML) situation where the true
semantics are imperfectly recognized.

Finally, a comprehensive overhaul of the XSLT processing was
carried out which avoids the divergence between
generation of the various HTML family of markup.
The stylesheets are highly parameterized so that they
are both more general, and yet allow generation of HTML5 specific markup;
they should allow extension to further HTML-like applications like ePub.
Command-line options make these parameters available to the user.

While the stylesheets are much more consistent and modular,
allowing easy extension and customization,
other changes lessen the need to customize.
The set of CSS class names have been made much more consistent
and predictable, if somewhat verbose, so that it should be easier
for users to style the generated HTML as they wish.
Additionally, a \texttt{resource} element has been defined
which allows binding developers to request certain CSS
or JavaScript files or fragments to be associated with the document.
A converted AMS article, now finally looks (somewhat) like
an AMS article! 

\subsection{Unification}\label{sec:unification}
Although the separation of the conversion and post-processing phases
is a natural one from the developer's document processing point of view,
it is sometimes artificial to users. Moreover, keeping
the phases too far separated inhibits interesting applications,
such as envisioned by the Daemon (see section \ref{sec:daemon})
and automated document processing systems such as the one used for arXMLiv.
Thus, we have undertaken to bring all processing back
under a single, consistent, umbrella, whether running in command-line mode,
or in client/server mode.  The goal is to simplify the common use-case
of converting a single document to HTML, while still enabling
the injection of intermediate processing.

Some steps in that direction include more consistent
error reporting at all phases of processing, with
embedded `locator' information so that the original source
of an error can (usually) be located in the source.
Additionally, logs include the current SVN revision number
to better enable tracking and fixing bugs.

\section{Daemon Experimental Branch}\label{sec:daemon}
The Daemon branch \cite{latexml:branch:online} hosts experimental developments,
primarily the development of client/server modules that
support web services, optimize processing and improve the integration with external applications. Since the last report in CICM's S\&P track \cite{GinStaKoh:latexmldaemon11}, the focus has fallen on increasing usability, security and robustness.

The daemonized processing matured into a pair of robust HTTP servers, one optimized for local batch conversion jobs, the other for a real-time web-service, and a turnkey client executable that incorporates all shapes and sizes of \LaTeXML\ processing. Showing a commitment to maintaining prominent conversion scenarios, shorthand user-defined \texttt{profiles} were introduced in order to simplify complex \LaTeXML\ configurations, e.g. those of sTeX and PlanetMath\cite{planetmath}. An internal redesign of the configuration setup and option handling of \LaTeXML\ contributed to facilitating these changes and promises a consistent internal API for supporting both the core and post-processing conversion phases.

The RESTful \cite{fielding00} web service offered via the Mojolicious \cite{Mojolicious} web framework now also supports multi-file {\LaTeX} manuscripts via a ZIP archive workflow, also facilitated by an upload interface. Furthermore, the built-in web editor and showcase \cite{latexml-showcase} is available through a websocket route and enjoys an expanded list of examples, such as a {\LaTeX} Turing machine and a PSTricks graphic. 

A significant new experimental feature is the addition of an ambiguous grammar for mathematical formulas. Based on Marpa \cite{Marpa}, an efficient Earley-style parser, the grammar embraces the common cases of ambiguity in mathematical expressions, e.g. that induced by invisible operators and overloaded operator symbols, in an attempt to set the stage for disambiguation to a correct operator tree. The current grammar in the main development trunk is heuristically geared to unambiguously recognize the mathematical formulas commonly used in DLMF and parts of arXiv. The long-term goal is for the ambiguous grammar to meet parity in coverage and implement advanced semantic techniques in order to establish the correct operator trees in a large variety of scientific domains.

It is anticipated that the bulk of these developments
will be merged back into the main trunk for the 0.8 release. The new ambiguous grammar and Mojolicious web service are two notable exceptions, which will not make master prior to the 0.9 release.

\section{Outlook}
Although development was never stagnated, an official release is long overdue;
a \LaTeXML\ 0.8 release is planned for mid-2013.
It will incorporate the enhancements presented here:
support for several \LaTeX\ graphics packages, such as Tikz and Xypic;
an overhauled XSLT and CSS styling framework;
and a merge of daemonized processing to the master branch.

\printbibliography

\end{document}